\documentclass[prd,aps]{revtex4}

\begin{document}
\preprint{{hep-th/yymmnnn} \hfill {UCVFC-DF-17-2005}}
\title{Electromagnetism and multiple-valued loop-dependent wave functionals}
\author{Lorenzo Leal $^{1,2}$}

\affiliation {  1.
Grupo de Campos y Part\'{\i}culas, Centro de  F\'{\i}sica Te\'orica y Computacional,
Facultad de Ciencias, Universidad Central de Venezuela, AP 47270,
Caracas 1041-A, Venezuela. \\ 2. Departamento de F\'{\i}sica, Universidad Sim\'on
Bol\'{\i}var,\\ Aptdo. 89000, Caracas 1080-A, Venezuela.\\}

\begin{abstract}
We quantize the Maxwell theory in the presence of a electric charge in a "dual" Loop Representation, i.e. a geometric representation
of magnetic Faraday's lines. It is found that the theory can be seen as a theory without sources, except by the fact that the wave functional becomes  multivalued. This can be seen as the dual counterpart of what occurs in Maxwell theory with a magnetic pole, when it is quantized in the ordinary Loop Representation. The multivaluedness can be seen as a result of the multiply-connectedness of the configuration space of the quantum theory.

\end{abstract}

\maketitle

It is well understood that in ordinary quantum mechanics in multiply connected configuration spaces,  wave functions may be allowed to be multivalued \cite{morandi,balachandran,balachandran2}. In those cases, multivaluedness is manifested through a phase factor in the wave function. This factor carries a representation of the fundamental group of the configuration space.
A celebrated example of this feature is the quantum model of two non-relativistic point particles in the plane interacting trough the Chern-Simons field \cite{leinaas,wilczek,iengo}. The relative motion results to be described by the Hamiltonian

\begin{equation}\label{uno}
H=\frac{(\overrightarrow{P}- e \overrightarrow{A})^{2}}{2m},
\end{equation}
with

\begin{equation}\label{dos}
A_{i}=\frac{k\, \varepsilon_{ij} \,r ^{j}}{\mid \overrightarrow{r}\mid ^{2}},
\end{equation}
where $r ^{j}$ are the components of the relative position vector. Since the potential vector can be written as $\overrightarrow{A}= k \overrightarrow{\nabla} \varphi $, $\varphi$ being the angle in polar coordinates, the Hamiltonian can be finally written down as

\begin{equation}\label{tres}
H=\frac{(\overrightarrow{P})^{2}}{2m},
\end{equation}
provided that the ordinary wave functional $\psi (\overrightarrow{r})$ be replaced by a multivalued wave functional $\psi ( n,\overrightarrow{r})= \exp (iken) \, \psi (\overrightarrow{r})$, where n is an integer that measures how many turns around the origin the vector $\overrightarrow{r}$ has taken. Hence, the topological interaction becomes translated into unusual boundary conditions of the wave function.

In reference \cite{leal} it was discussed an example of how this mechanism may also be realized in the framework of quantum field theory. Concretely, in that reference it was studied the Maxwell field in the presence of a static Dirac's magnetic monopole \cite{dirac,dirac2}, within the Loop Representation formalism \cite{dibartolo,gambini1,gambini2}. It was shown that the quantum theory may be casted in the form of a free theory, except by the fact that the loop-dependent wave functional carries a non-trivial one-dimensional unitary representation of the fundamental group associated to the space of loops in $R^{3}-{origin}$. This phase factors takes into account the presence of the monopole in the origin.
The multivaluedness of the loop-dependent wave functional is attained by means of the following construction. The loop space is "enlarged" to a space of open surfaces. Then it is shown that  the wave functional depends on these surfaces  through a topological phase factor, the phase being proportional to the solid angle subtended by the loop that bounds the surface, measured with respect  to the monopole. Hence, the multivaluedness is measured in terms of the number of times that the surface wraps around the monopole.

In this note, we shall provide the "dual" example of the above case, which, as will be seen, results to be easier to obtain than the former one. We shall consider the Maxwell field in the presence of a static electric charge in the origin, and we will quantize it in a geometric representation dual to the ordinary Loop Representation.

Let us begin by summarizing the results of reference \cite{leal}. There, the starting point is the first-order Schwinger action \cite{schwinger}

\begin{equation}\label{cuatro}
S=\int dx^{4} \left (A_{\mu} J^{\mu}_e+B_{\mu} J^{\mu}_m-\frac
{1}{2}F^{\mu\nu}(\partial_{\mu}A_{\nu}-\partial_{\nu}A_{\mu})+\frac
{1}{4}F^{\mu\nu}F_{\mu\nu}\right ),
\end{equation}
where $B$ is given by
\begin{eqnarray}\label{cinco}
B_{\mu}(x)&=& \int dy^{4}
*F_{\mu\nu}(y)f^{\nu}(y-x)+\partial_{\mu}\lambda(x).
\end{eqnarray}
Here, $\lambda$ is an arbitrary function. The function  $f$ must obey
\begin{eqnarray}
\partial_{\mu}f^{\mu}(y)&=&\delta^{4}({y}),\label{seis}
\end{eqnarray}
and can be taken as

\begin{equation}\label{f}
f^{\nu}(y)=-\frac{1}{4\pi}\frac{y^{i}}{\vec{y}\,^3}\delta^{\nu}_{i}
\delta(y_0).
\end{equation}
The Hodge dual $*F$ is given by $*F_{\mu\nu}=\frac{1}{2}\epsilon_{\mu\nu\alpha\beta}F^{\alpha\beta},$
 and $J_e$ ($J_m$) denote the electric
(magnetic) current density. In (\ref{dos}) the independent fields are taken to be  $A_{\nu}$ and $F_{\mu\nu}$.

Since the source considered is just a static monopole at the origin,  we take $J_e=0$ in equation (\ref{dos}) and
\begin{equation}
J^{\mu}_m(x)=g\delta^{\mu}_0 \delta^{3}(\vec{x}).
\end{equation}

The canonical quantization a la Dirac \cite{dirac3} yields the following results \cite{leal}. After the resolution of the second class constraints that rise in the canonical formalism, by means of the Dirac brackets construction, one is left with a pair of canonical variables, $\hat{A}_i(x)$ and $\hat{\Pi^j}(y)$, obeying the equal time commutators

\begin{eqnarray}
\label{algebra1} [\hat{A}_i(x),\hat{A}_j(y)]&=& 0,\\
\label{algebra2} \left[\hat{\Pi}^i(x),\hat{\Pi}^j(y)\right]&=& 0,\\
\label{algebra3} \left[ \hat{A}_i(x),\hat{\Pi^j}(y)\right]&=&
i\delta^{j}_i\delta^{3}(\vec x-\vec y).
\end{eqnarray}
 The first class constraints define the physical states $|\Psi>$ as
those that satisfy
\begin{eqnarray}\label{c1}
\partial_i\hat{\Pi}^i(x)|\Psi>&=& 0.
\end{eqnarray}
On the physical subspace, the dynamics is given by the
Schr\"{o}dinger equation
\begin{equation} i\partial_t|\Psi_t>=\hat{H}|\Psi_t>,
\end{equation}
with
\begin{equation}\label{phys}
\hat{H}=\int d\vec x\,^{3}\left[\frac{1}{2}\hat{\Pi}
_{i}^2+\frac{1}{4}\left(\partial_i \hat{A}_j-\partial_j
\hat{A}_i-b_{ij}\right)^{2}\right].
\end{equation}
Thus, we obtain that the static monopole manifests in the theory
just through the external field $b_{ij}$, that must be subtracted
from the curl of the vector potential to give the magnetic field
operator.
In the Loop Representation \cite{dibartolo}, the electric field (which is $\hat{\Pi}^i(x)$) becomes the form factor of the Loop $C$

\begin{eqnarray}\label{E}
E^{i} (\overrightarrow{x}) \psi [C] = e T^{i} (\overrightarrow{x},C) \psi [C] .
\end{eqnarray}
with
\begin{eqnarray}\label{E}
T^{i} (\overrightarrow{x},C) = \oint dy^{i} \delta ^{3} (\overrightarrow{x}-\overrightarrow{y}).
\end{eqnarray}
On the other hand, the curl of the potential (which is the magnetic field in the absence of magnetic poles) is given by the loop derivative \cite{dibartolo}

\begin{eqnarray}\label{B}
F_{ij}(\overrightarrow{x}) \psi [C] = \frac{i}{e} \triangle _{ij} (\overrightarrow{x}) \psi [C] .
\end{eqnarray}
These realizations fulfil the canonical algebra and the Gauss law. Then, the Hamiltonian in the Loop Representation is given by

\begin{eqnarray}\label{HLR}
H= \int dx^{3} (\frac{1}{2} e^{2} (T^{i} (\overrightarrow{x},C)^{2}) - \frac{1}{4 e^{2}} (i \triangle _{ij} (\overrightarrow{x}) - e b _{ij} (\overrightarrow{x}))^{2}).
\end{eqnarray}
Now, in reference \cite{leal} we showed that  the term $b_{ij}$ can be "dropped" from the Hamiltonian provided that instead of dealing with "ordinary" loop-dependent wave functionals $\psi [C]$, we admit multivalued loop-dependent wave functionals $\psi [C, n]= \exp (iegn)\,\Psi[C]$. Here, as in the former Chern-Simons example, $n$ is an integer that counts how many times the closed surface that a loop sweeps adiabatically,  wraps around the monopole (i.e. around the origin of $R^{3}$). Again, the interaction is turned into non usual boundary conditions of the wave functional.

Now we turn to consider the Maxwell field in the presence of a static point electric charge at the origin. Whether we start from the Schwinger first order action or from the usual second order Maxwell action, the canonical formulation yields the same canonical commutators as before, and the Hamiltonian is

\begin{equation}\label{phys}
\hat{H}=\int d\vec x\,^{3}\left[\frac{1}{2}\hat{\Pi}
_{i}^2+\frac{1}{4}\left(\partial_i \hat{A}_j-\partial_j
\hat{A}_i\right)^{2}\right],
\end{equation}
which also corresponds to the Hamiltonian for the case without sources. The electric charge manifests this time in the Gauss constraint

\begin{eqnarray}\label{c1}
(\partial_i\hat{\Pi}^i(\overrightarrow{x} )- e \delta ^{3}(\overrightarrow{x}))|\Psi>&=& 0.
\end{eqnarray}

We are interested in realizing the canonical algebra in a geometric representation dual to the "electric" Loop Representation. To this end, we recall briefly the space of surfaces framework \cite{pio1,pio}. Take piecewise smooth oriented surfaces $\Sigma$ in $R^{3}$.  Then consider the "surface form-factor"
\begin{equation}\label{2.16}
T^{ij}(\vec{x},\Sigma)=\int
d\Sigma^{ij}_y\,\delta^{(3)}(\vec{x}-\vec{y}).
\end{equation}
Here $d\Sigma^{ij}_y$ is the surface element
\begin{equation}\label{2.18}
d\Sigma^{ij}_y=(\frac{\partial y^i }{\partial s}\frac{\partial
y^j}{\partial r} - \frac{\partial y^i}{\partial r}\frac{\partial
y^j}{\partial s})ds dr,
\end{equation}
with $s,r$ being surface parameters.
Instead of loop-dependent functionals, one can consider surface-dependent ones
$\Psi[\Sigma]$ and introduce the surface derivative
$\delta_{ij}(\vec x)$, which measures how
$\Psi[\Sigma]$ changes when an element of surface of infinitesimal area
is $\sigma_{ij}$ is appended to  $\Sigma$  at the point $\overrightarrow{x}$, up to first order in
$\sigma_{ij}$ :
\begin{equation}\label{2.17}
\Psi [\delta\Sigma\cdot\Sigma]= (1 + \sigma^{ij}\delta_{ij}(\vec
x))\Psi [\Sigma].
\end{equation}
The surface-derivative of the form factor is given by
\begin{equation}\label{2.22}
\delta_{ij} (\vec{x}) T^{kl}(\vec{y},\Sigma)=\frac
12\left(\delta_{i}^{k}\delta_{j}^{l} -\delta_{i}^{l}
\delta_{j}^{k}\right)\delta^{(3)}(\vec{x}-\vec{y}).
\end{equation}
In virtue of this equation, the canonical algebra can be realized in the space of surface-dependent functionals as
\begin{equation}\label{2.23}
\hat{A}_{i}(\vec{x}) = g \varepsilon_{ijk} T^{jk}(\vec{x},\Sigma) ,
\end{equation}

\begin{equation}\label{2.24}
\hat{E}^{i}(\vec{x}) = \frac{-i}{g}\,\varepsilon^{ijk} \delta_{jk}(\vec{x}) ,
\end{equation}
where $g$ is a constant that sets the strength of a unit of magnetic flux: the curl of $\hat{A}_{i}(\vec{x})$, which is the magnetic field, is given by $\hat{B}^{i}(\vec{x})= g T^{i}(\vec{x},\partial \Sigma)$. Hence, $g$ measures the amount of magnetic flux carried by a "magnetic Faraday's line" along the boundary of the surface $\partial \Sigma$, whose form factor is $T^{i}(\vec{x},\partial \Sigma)$.
In the absence of electric charges the Gauss constraint (\ref{c1}) forces the wave functional to depend only on the boundary $\partial \Sigma$ of the surface. This can be readily seen by noticing that the divergence of the electric field becomes, in the surface space, the "die" derivative, i.e., a  derivative that measures how the surface-dependent functional changes when  an infinitesimal closed surface is appended to its argument. If the die derivative vanishes, the functional depends only on boundaries of surfaces (loops), inasmuch the vanishing of the loop derivative says that functional depends only on boundaries of loops, i.e., points. Thus, we obtain the dual representation of the free Maxwell Loop Representation: instead of being lines of electric field, the dual loops are lines of magnetic field. In turn, the electric field is the "magnetic loop derivative" in this representation.

Now, let us turn back to the case when there is a point charge in the origin of space. The Gauss constraint (\ref{c1}), in the surface representation may be written down as
\begin{eqnarray}\label{2.25}
( \frac{-i}{g}\, \varepsilon^{ijk} \partial_i  \delta_{jk}(\vec{x}) - e \delta ^{3}(\vec{x})) \Psi[\Sigma] &=& 0.
\end{eqnarray}
Without lose of generality, we can set $\Psi[\Sigma] = \exp(if[\Sigma]) \Phi[\Sigma]$. Substituting this expression in (\ref{2.25}), we find that $\Phi[\Sigma]$
obeys the free (i.e. without charges) Gauss law, provided that $f[\Sigma]= \frac{eg}{4\pi} \Omega (\Sigma)$, where $\Omega (\Sigma)$ is the solid angle subtended by the surface as seen from the origin. Hence the physical sector of the surface-dependent wave functionals is given by
\begin{eqnarray}\label{2.26}
\Psi[\Sigma] = \exp(i \frac{eg}{4\pi} \Omega (\Sigma)) \Phi[\delta \Sigma].
\end{eqnarray}
But once again, we see that the dependence of the phase factor in the surface $\Sigma$ is of topological nature: it depends on the solid angle $\Omega (\Sigma)$ subtended by the surface as seen from the electric charge. Given the loop $\delta \Sigma$ that bounds the surface, this solid angle is fixed except by integer multiples of $4\pi$. Hence, equation (\ref{2.26}) can be interpreted as follows: the physical sector of the theory (i.e., that which satisfies the Gauss law constraint (\ref{2.25}) is given by multiple valued loop-dependent functionals, the multivaluedness given as a topological phase factor that sees how many times the loop winds around the electric charge. Thus it can be concluded that in the dual Loop Representation (i.e., a polymer-like representation of magnetic Faraday's lines, in the terminology of references \cite{Madhavan,Madhavan2}) of electromagnetism with point electric charges, the effect of the charges can be encoded into unusual boundary conditions that the loop-dependent wave functionals must obey. The canonical algebra, and the Hamiltonian, on the other and, are the same as in the free case.

 As in the theory with magnetic poles \cite{leal}), this fact can be seen as a generalization of what happens
in ordinary quantum mechanics in multiply connected configuration
spaces \cite{morandi,balachandran,balachandran2}. In those  cases,
the wave function is allowed to be multivalued due to the multiple connectedness of the configuration space.
In the present formulation, the configuration space is the
space of loops  in $R^{3}- \{origin\}$. Hence a ''point'' in the space should be taken as
a magnetic loop , while the ''closed curves'' swept by those loops loops become closed
surfaces. The topological phase factor appearing in the wave functional, which classifies the
surfaces according to the manner they  wrap the electric charge, is precisely a one-dimensional
representation of the fundamental group of the configuration space, in agreement with what occurs both in the case of electromagnetism with a magnetic monopole, and in the ordinary quantum mechanical case in multiply-connected configuration spaces.

The author acknowledges fruitful discussions with Frank Vera.


\begin{thebibliography}{99}


\bibitem{morandi}
G. Morandi, The Role of Topology in Classical and Quantum Physics
( Springer-Verlag, 1992)

\bibitem{balachandran}
A.P.Balachandran, G.Marmo, B.S.Skagerstam and A.Stern, Classical
Topology and Quantum States (World Scientific, 1991)

\bibitem{balachandran2}
A.P.Balachandran, Found. Phys. \textbf{24}, 455 (1994).

\bibitem{leinaas}
J.M. Leinaas and J. Myrheim, Il Nuov.Cim. \textbf{37B}, 1 (1977).

\bibitem{wilczek}
F. Wilczek, Phys. Rev. Lett. \textbf{49}, 957 (1982).

\bibitem{iengo}
R.Iengo and K.Lechner, Phys.Rep. \textbf{213 4}, 180 (1992).

\bibitem{leal}
L.Leal and A.Lopez, Journ.Math.Phys \textbf{47,1}, 230601 (2005).

\bibitem{dirac}
P.A. M Dirac, Proc.\ Roy.\ Soc.\ Lond. {\bf A 113}, 60, (1931)
\bibitem{dirac2}
P.A.M. Dirac, Phys.\ Rev.\ {\bf 74 }, 317, (1948)


\bibitem{dibartolo}
C. di Bartolo, F. Nori, R. Gambini and A. Tr\'{i}as,\ Nuovo\
Cimento Soc. Ital. Fis. {\bf 38}, 497, (1983)

\bibitem{gambini1}
R. Gambini and A. Tr\'{i}as, Phys.\ Rev.\ D {\bf 27}, 2935, (1983)

\bibitem{gambini2}
R. Gambini and J. Pullin, Loops Knots, Gauge Theory and Quantum
Gravity (Cambridge University Press, 1996)

\bibitem{schwinger}
J. Schwinger, Phys.\ Rev.\ {\bf D12 }, 3105, (1975)

\bibitem{dirac3}
P.A.M. Dirac, Lectures on quantum mechanics (Yeshiva University,
New York,1964)


\bibitem{pio1} P.~J.~Arias, C.~Di Bartolo, X.~Fustero, R.~Gambini and A.~Tr\'{i}as,
Int.\ J.\ Mod.\ Phys.\ {\bf A7}, 737 (1992)

\bibitem{pio}
P.J.Arias, E.Fuenmayor and L.Leal, Phys.\ Rev.\ D {\bf 69 },
125010 (2004)

\bibitem{Madhavan}M. Varadarajan, Phys.Rev. {\bf D61}, 104001 (2000).

\bibitem{Madhavan2}M. Varadarajan, Phys.Rev.
\textbf{D64} 104003 (2001).




\end{thebibliography}
\end{document}